\documentclass[english,twocolumn,superscriptaddress, aps, prd,longbibliography]{revtex4-1}
\usepackage[T1]{fontenc}
\usepackage[utf8]{inputenc}
\usepackage[active]{srcltx}
\usepackage{babel}
\usepackage{amsmath}
\usepackage{graphicx}
\usepackage[unicode=true,pdfusetitle,
 bookmarks=true,bookmarksnumbered=false,bookmarksopen=false,
 breaklinks=false,pdfborder={0 0 1},backref=false,colorlinks=false]
 {hyperref}

\makeatletter

\providecommand{\tabularnewline}{\\}

\usepackage[printonlyused]{acronym}

\AtBeginDocument{%
    \newwrite\bibnotes
    \def\bibnotesext{Notes.bib}
    \immediate\openout\bibnotes=\jobname\bibnotesext
    \immediate\write\bibnotes{@CONTROL{REVTEX41Control}}
    \immediate\write\bibnotes{@CONTROL{%
    apsrev41Control,author="08",editor="1",pages="1",title="0",year="1"}}
     \if@filesw
     \immediate\write\@auxout{\string\citation{apsrev41Control}}%
    \fi
}%

\makeatother

\begin{document}
\title{Compensation of front-end and modulation delays in phase and ranging
measurements for time-delay interferometry}
\author{Philipp Euringer}
\email{philipp.euringer@airbus.com}

\affiliation{Airbus Space Systems, Airbus Defence and Space GmbH, Claude-Dornier-Straße,
88090 Immenstaad am Bodensee, Germany}
\author{Niklas Houba}
\altaffiliation{Present address: Institute of Geophysics, Department of Earth and Planetary Sciences, ETH Zurich, Sonneggstr. 5, 8092 Zurich, Switzerland}

\affiliation{Airbus Space Systems, Airbus Defence and Space GmbH, Claude-Dornier-Straße,
88090 Immenstaad am Bodensee, Germany}
\author{Gerald Hechenblaikner}
\author{Oliver Mandel}
\affiliation{Airbus Space Systems, Airbus Defence and Space GmbH, Claude-Dornier-Straße,
88090 Immenstaad am Bodensee, Germany}
\author{Francis Soualle}
\affiliation{Airbus Space Systems, Airbus Defence and Space GmbH, Willy-Messerschmitt-Straße
1, 82024 Taufkirchen, Germany}
\author{Walter Fichter}
\affiliation{Institute of Flight Mechanics and Controls, University of Stuttgart,
Pfaffenwaldring 27, 70569 Stuttgart, Germany}
\begin{abstract}
In the context of the Laser Interferometer Space Antenna (LISA), the
laser subsystems exhibit frequency fluctuations that introduce significant
levels of noise into the measurements, surpassing the gravitational
wave signal by several orders of magnitude. Mitigation is achieved
by means of time-shifting individual measurements in a data processing
step known as time-delay interferometry (TDI). The suppression performance
of TDI relies on accurate knowledge and consideration of the delays
experienced by the interfering lasers. While considerable efforts
have been dedicated to the accurate determination of inter-spacecraft
ranging delays, the sources for delays onboard the spacecraft have
been either neglected during TDI processing or assumed to be known.
Contrary to these assumptions, analog delays of the phasemeter front
end and the laser modulator are not only large but also prone to change
with temperature and heterodyne frequency. This motivates our proposal
for a novel method enabling a calibration of these delays on-ground
and in-space, based on minimal functional additions to the receiver
architecture. Specifically, we establish a set of calibration measurements
and elucidate how these measurements are utilized in data processing,
leading to the mitigation of the delays in the TDI Michelson variables.
Following a performance analysis of the calibration measurements,
the proposed calibration scheme is assessed through numerical simulations.
We find that in the absence of the calibration scheme, the assumed
drifts of the analog delays increase residual laser noise at high
frequencies of the LISA measurement band. A single, on-ground calibration
of the analog delays leads to an improvement by roughly one order
of magnitude, while re-calibration in space may improve performance
by yet another order of magnitude. Towards lower frequencies, ranging
error is always found to be the limiting factor for which countermeasures
are discussed.
\end{abstract}
\maketitle

\section{Introduction\label{sec:Introduction}}

During the past years several space missions aiming to detect gravitational
waves have undergone intense study activities and are expected to
soon enter the implementation phase. Among these missions are LISA
\cite{amaro2017laser}, a joint ESA-NASA mission, the Chinese TAIJI
\cite{luo2020taiji} and TianQin \cite{luo2016tianqin} missions,
and the Japanese DECIGO mission \cite{sato2017status}. The LISA constellation
comprises three spacecraft (S/C) which form an approximately equilateral
triangle of 2.5 million km arm-length and rotate around their common
center in an Earth trailing orbit. Gravitational waves are detected
through interferometric measurements of the relative distance changes
between two freely floating test masses along each arm of the constellation,
where each S/C transmits to and receives light from both its neighboring
S/C \cite{amaro2017laser}. The measurement for each arm is broken
down into individual measurements of test-mass position relative to
an optical bench (OB) and OB of one S/C relative to the OB of the
opposite S/C, based on a concept referred to as ``strap-down interferometry''
\cite{Gath_2009_LISA_Mission}. Given that the strain sensitivity
of the detector is proportional to the distance between S/C, attaining
the targeted strain sensitivity of $10^{-21}$ within the 0.1 mHz
to 1 Hz measurement frequency range requires the LISA constellation
to extend across several million kilometers.\\
Owing to seasonally changing gravitational disturbances, the arm-lengths
oscillate around their mean values with amplitudes of several thousand
kilometers and relative velocities of around 15 m/s \cite{OttoThesis2015},
leading to slowly varying Doppler shifts of the received laser frequency
on the order of several megahertz. To ensure that the interferometric
beat-note frequency (heterodyne frequency) remains within the detection
bandwidth of the phasemeter, the lasers of all 6 optical links are
frequency offset-locked relative to one another and require occasional
changes of the frequency offset to account for the slow evolution
of the heterodyne frequency over time, as defined in a ``frequency
offset plan'' similar to the one defined for TAIJI \cite{zhang2022inter}.
Despite using highly stable laser sources, the associated laser frequency
noise transforms into large interferometric phase fluctuations across
the arm-length which spoil the measurement performance. This physical
show-stopper for long-distance heterodyne interferometry is overcome
by a post-processing technique referred to as time-delay interferometry
(TDI) \cite{tinto2021time,Vine2010TDI}. In TDI, laser noise is suppressed
by time-shifting and linearly combining the interferometric measurements.
The goal is to obtain synthetic measurement observables representing
virtual equal arm-length interferometers. While there are many variants
of TDI observables, each having their distinct advantages for gravitational
wave analysis and instrument diagnostic, the $X,~Y,~Z$ observables,
representing synthetic Michelson interferometers, are among the most
frequently used ones in LISA science data processing research, and
we will also adopt them in this paper. The Michelson variables allow
completely suppressing laser frequency noise for unequal arm-lengths
of a static constellation in TDI first generation (TDI 1.0)  \cite{Tinto2002}.
However, accounting for changing arm-lengths and a rotating constellation
required developing TDI second generation (TDI 2.0), where more complicated
linear combinations of interferometric measurements are built to suppress
laser noise to the required level. Both, TDI 1.0 and TDI 2.0 require
the arm-lengths to be known so that the individual interferometric
measurements can be shifted in time correctly. Consequently, any error
in the arm-length estimate decreases the efficiency of the TDI suppression
performance.

In order to supply TDI with the required inter S/C distance, referred
to as ``range'', auxiliary measurements are continuously performed
for LISA \cite{heinzel2011auxiliary}. These measurements rely on
modulating the carrier phase of the transmitted laser beam with a
pseudo-random noise (PRN) code sequence which is correlated with a
local replica on the receiving S/C in order to determine the pseudo-range
within the ambiguity range defined by the period of the PRN sequence
in space ($\approx300\,$km), which is a similar approach as the one
used for GPS navigation systems in the radio-frequency regime. Regular
coarse S/C position measurements from ground combined with orbit prediction
of the S/C trajectories, allow for accurately resolving the ambiguity
and determining the absolute ranges between S/C.

Note that for this scheme to work it is additionally required that
the PRN code sequences modulated onto the transmitted beam on one
S/C are synchronized to those of the local replica on the other S/C,
which is the case if the clocks of all 3 S/C in the LISA constellation
are synchronized to a common reference time. Ranging errors are closely
related to clock synchronization errors, both leading to an offset
in the alignment of individual measurements in the synthesis of TDI
observables. Therefore, clock synchronization is generally required
for synthesis of the virtual equal arm-length interferometers in TDI
and performed as one of the post-processing steps in the LISA Initial
Noise Reduction Pipeline (INReP) \cite{OttoThesis2015}, although
there has been a recent proposal for a TDI variant without the need
for clock synchronization \cite{Hartwig2022}.

Previous post-processing models have neglected or assumed perfect
knowledge \cite{reinhardt2023ranging} of the delays in the phase
measurement chain, including delays caused by the analog phasemeter
front end and digital delays in the signal processing of the phasemeter
back end, affecting not only the 6 long-arm interferometers measuring
the inter-S/C distances but all 18 interferometers used in the LISA
constellation. Additionally, the 6 auxiliary ranging measurements
of the inter-S/C distance are affected by delays of the laser PRN
modulation with respect to the S/C clock on the transmitter side as
well as phasemeter front-end and back-end delays on the receiver side.

In this paper, we focus on these novel aspects and investigate how
these delays can be accurately calibrated on ground and re-calibrated
in space by only small functional additions to the current baseline
of the receiver architecture. After identifying these delays in section
\ref{subsec:Ranging-Biases}, we define a set of calibration measurements
in section \ref{subsec:Calibration-Measurement}. In section \ref{sec:TDI-Processing},
we show how associated measurements are applied in a post-processing
step and in TDI resulting in a compensation of these delays in the
TDI Michelson variables. In section \ref{sec:Calibration-performance},
we assess the calibration and ranging performance due to sources of
random noise, based on current baseline models of the LISA receiver
architecture and physical signal parameters, and account for the errors
in the simulations presented in section \ref{sec:Simulation-and-Results}.
Within these numerical simulations, the performance impact on TDI
is assessed for 3 major cases: (1) the delays are not compensated,
(2) the delays are only compensated based on calibration measurements
performed on ground, and (3) the delays are compensated based on periodic
calibration measurements performed in space. For case (2) the frequency
noise suppression is increased by roughly one order of magnitude with
respect to case (1) for higher measurement frequencies from 10 mHz
to 1 Hz, while for case (3) the performance improves by roughly one
order of magnitude with respect to case (2) for higher measurement
frequencies from 100 mHz to 1 Hz. These findings clearly emphasize
the importance of accurate delay calibration and adjustment of calibration
parameters during the operational lifetime of LISA. At low frequencies
of the LISA measurement band, performance is limited by the ranging
error and possibilities are discussed to greatly improve the ranging
performance by additional small modifications to the baseline of the
receiver architecture.

\section{Measurement Principle and Delays\label{subsec:Ranging-Biases}}

\begin{figure}[h]
\centering{}\includegraphics[width=.6\linewidth]{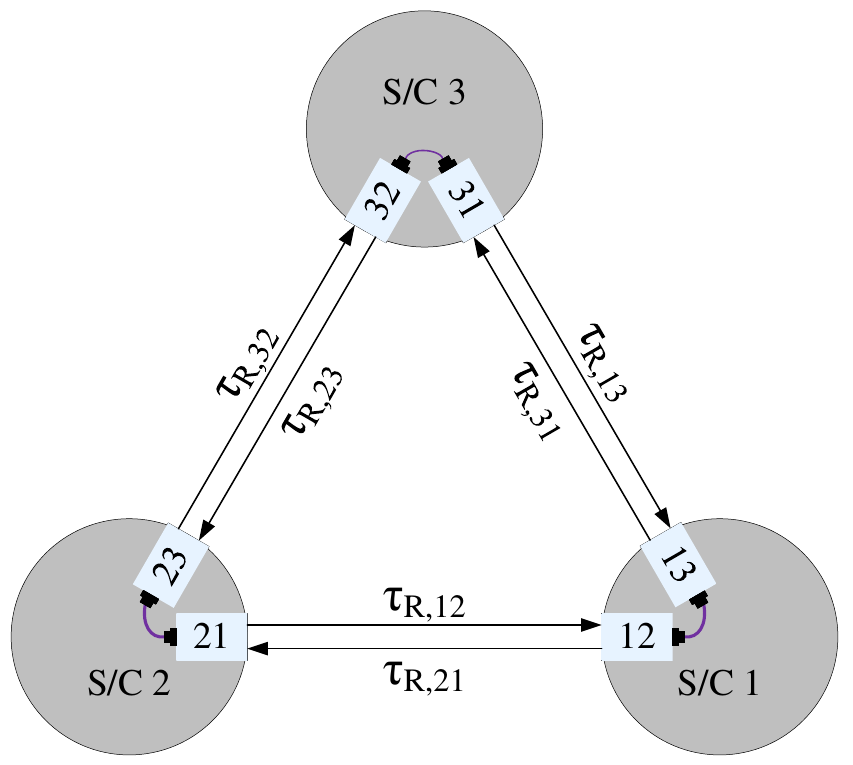}\caption{The LISA constellation consists of three S/Cs each hosting two OBs,
highlighted in light blue. The light travel times among these OBs
are indicated via the parameter $\tau_{\text{R}}$. \label{fig:LISA-constellation}}
\end{figure}
\begin{figure}[h]
\centering{}\includegraphics[width=1\linewidth]{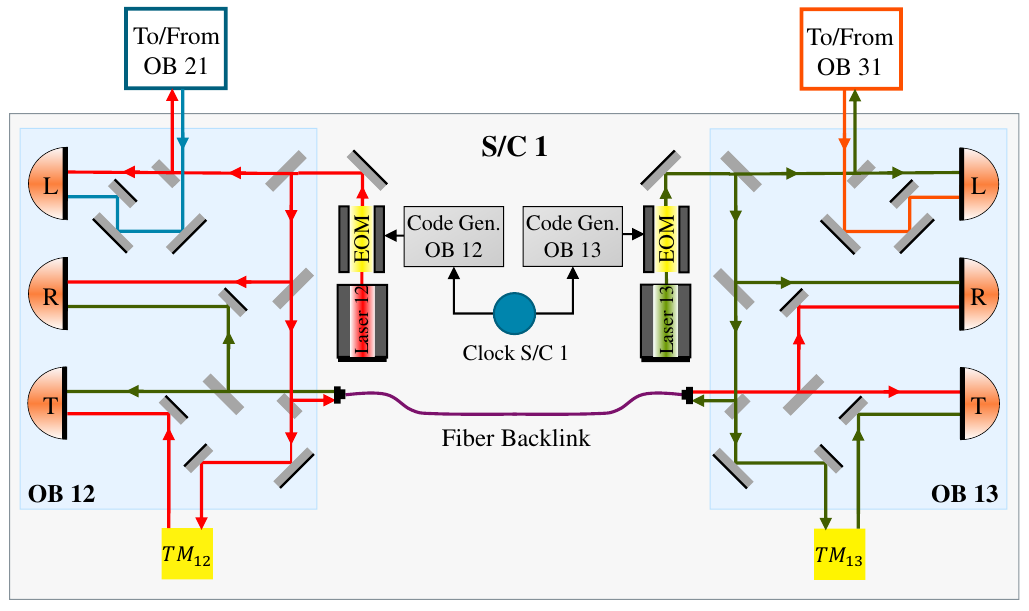}\caption{A simplified sketch indicating the interferometers present at S/C
1 is depicted. The S/C hosts two OBs. Each OB is equipped with three
interferometers: the long-arm interferometer (L), the reference interferometer
(R), and the test-mass interferometer (T). The photodiodes of these
interferometers have been labeled with the respective letter. The
OBs are connected via a fiber backlink highlighted in purple. For
the sake of clarity, the beams of each laser have been distinctly
color-coded. At the EOM lasers are modulated with a chip sequence
supplied by the code generation unit, indicated via ``Code Gen.''.
\label{fig:Identification-of-interferometer}}
\end{figure}

Figure \ref{fig:LISA-constellation} depicts a simplified schematic
of the LISA constellation. Each S/C hosts two OBs and two associated
free-falling test masses. The OBs will thereby be denoted as OB $ab$,
where $a$ represents the S/C the OB is located and $b$ the S/C the
laser of the OB is pointing to. In turn, each OB is equipped with
three interferometers, cf. Fig. \ref{fig:Identification-of-interferometer},
leading to 18 interferometers for the full constellation. The long-arm
interferometer (indicated via subscript L) measures relative distance
fluctuations among the local OB and the OB of the remote S/C. The
test-mass interferometer (indicated via subscript T) measures distance
changes of the free-falling test mass of the local OB relative to
the phase of the reference beam transmitted from the adjacent OB through
a backlink fiber. Finally, the reference interferometer (indicated
via subscript R) measures relative phase changes of the local reference
beam against the reference beam of the adjacent OB without the test
mass being part of the optical path.

\begin{figure*}
\centering{}\includegraphics[width=1\linewidth]{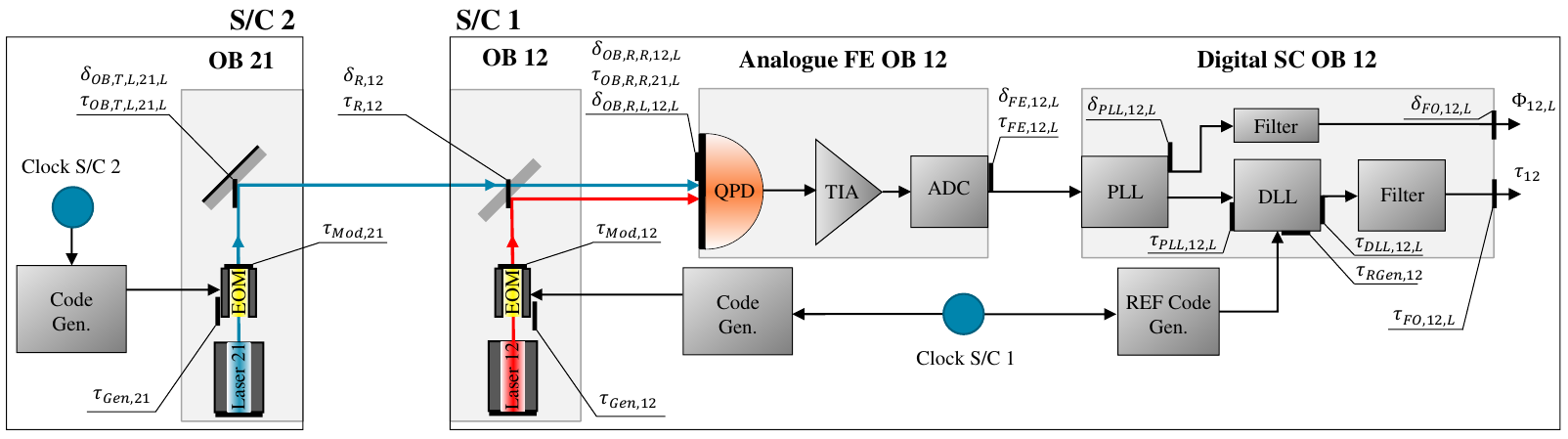}\caption{Schematic indicating the delays affecting the carrier phase and code
read-out at the long-arm interferometer of OB 12. Individual phase
and code delays are separated by thick black bars and labeled at the
end of the individual delays with symbols $\delta$ and $\tau$, respectively.\label{fig:Identification-of-ranging-bias}}
\end{figure*}

All of these 18 interferometers follow a common measurement principle,
depicted in Fig. \ref{fig:Identification-of-ranging-bias} \cite{danzmann2011LISA}.
Two laser beams with frequencies in the THz range ($\lambda=$ 1064
nm) are combined on an OB. The combined beams are processed by an
analog front end. Interference of the beams at a QPD (quadrature photodiode)
creates a beat note in the range of 6 to 24 MHz, depending on the
relative S/C motion \cite{EESA}. The resulting photocurrent serves
as input to a transimpedance amplifier (TIA), which converts the photocurrent
to a photovoltage. Thereby, the TIA is frequency selective, exhibiting
a bandpass in the range of 5 to 25 MHz \cite{Barranco2018}. Finally,
an ADC (analog-to-digital converter) circuit transforms the signal
to the digital domain, where the phase-readout is performed using
an all-digital phase-locked loop (PLL) \cite{Shaddock2006,EESA}.
\\
These processing stages unavoidably introduce delays in the interferometric
readouts, in addition to the delays arising from the individual routing
of the interfering beams. In Fig. \ref{fig:Identification-of-ranging-bias},
phase delays of the long-arm interferometer, denoted with the symbol
$\delta$, have been identified, similar to those described in \cite{reinhardt2023ranging}.
Each delay in Fig. \ref{fig:Identification-of-ranging-bias} is demarcated
by thick black lines, with corresponding labels provided at the end
of each delay.
\begin{table}
\centering{}\caption{Phase delays $\delta$ encompassed by the laser beams for OB $ab$
and \mbox{Interferometer $Z$} with $a,b\in\{1,2,3\}$, $a\protect\neq b$
and $Z\in\{\text{L},\text{T},\text{R}\}$.\label{tab:Front-End-Delays}}
\begin{tabular}{ll}
\hline 
\hline Interferometric phase delay & Symbol\tabularnewline
\hline 
OB delay at transmission (T) at local (L) S/C & $\delta_{\text{OB,T,L},ab,Z}$\tabularnewline
OB delay at reception (R) at local (L) S/C & $\delta_{\text{OB,R,L},ab,Z}$\tabularnewline
OB delay at reception (R) at remote (R) S/C & $\delta_{\text{OB,R,R},ab,Z}$\tabularnewline
Phase delay of analog front end & $\delta_{\text{FE},ab,Z}$\tabularnewline
PLL processing delay & $\delta_{\text{PLL},ab,Z}$\tabularnewline
Output filtering delay & $\delta_{\text{FO},ab,Z}$\tabularnewline
Ranging delay between S/C $b$ and S/C $a$ & $\delta_{\text{R},ab}$\tabularnewline
Phase delay due to test mass motion & $\delta_{\text{TM},ab}$\tabularnewline
Fiber backlink among OB $ba$ and OB $ab$ & $\delta_{\text{BL},ab}$\renewcommand{\tabularnewline}{\\ \hline}\tabularnewline
\hline 
\end{tabular}\renewcommand{\tabularnewline}{\\}
\end{table}
 The interferometric read-outs of the test-mass and reference interferometer
are delayed in a similar way except for the free space transmission
of the remote laser. The individual delays of all interferometers
are listed in Tab. \ref{tab:Front-End-Delays}. Equations \ref{eq:phi_L_0}
to \ref{eq:phi_R_0} state the phase measurement $\Phi(t)$ on OB
12 for the three interferometers and detail how these delays affect
the individual phase $\phi(t)$ of the interfering lasers. 
\begin{align}
\Phi_{12,L}(t & +\delta_{\text{FE},12,\text{L}}+\delta_{\text{PLL},12,\text{L}}+\delta_{\text{FO},12,\text{L}})\nonumber \\
 & =\phi_{21}(t-\delta_{\text{OB,T,L},21,\text{L}}-\delta_{\text{R},12}-\delta_{\text{OB,R,R},12,\text{L}})\nonumber \\
 & -\phi_{12}(t-\delta_{\text{OB,R,L},12,\text{L}}),\label{eq:phi_L_0}\\
\Phi_{12,T}(t & +\delta_{\text{FE},12,\text{T}}+\delta_{\text{PLL},12,\text{T}}+\delta_{\text{FO},12,\text{T}})\nonumber \\
 & =\phi_{13}(t-\delta_{\text{OB,T,L},13,\text{T}}-\delta_{\text{BL},12}-\delta_{\text{OB,R,R},12,\text{T}})\nonumber \\
 & -\phi_{12}(t-\delta_{\text{OB,R,L},12,\text{T}}-\delta_{\text{TM},12}),\label{eq:phi_T_0}\\
\Phi_{12,R}(t & +\delta_{\text{FE},12,\text{R}}+\delta_{\text{PLL},12,\text{R}}+\delta_{\text{FO},12,\text{R}})\nonumber \\
 & =\phi_{13}(t-\delta_{\text{OB,T,L},13,\text{R}}-\delta_{\text{BL},12}-\delta_{\text{OB,R,R},12,\text{R}})\nonumber \\
 & -\phi_{12}(t-\delta_{\text{OB,R,L},12,\text{R}}).\label{eq:phi_R_0}
\end{align}
The subscript of the phase measurement $\Phi(t)$ denotes the OB and
the interferometer, separated by a comma. Similarly, the subscript
of the laser phase $\phi$ denotes the OB the laser is associated
with. Note that the time argument of the phase delays $\delta$ has
been omitted in Eqs. \ref{eq:phi_L_0} to \ref{eq:phi_R_0} for readability.
Moreover, phase delays appearing equally in both interfering lasers
have been considered in the argument of the phase measurement $\Phi(t)$
. Interferometric readouts at the remaining OBs can be retrieved
through a cyclic permutation of the numerical indices.

Besides the interferometric measurement, the long-arm interferometers
allow for clock synchronization, absolute ranging, and data exchange
among the S/Cs (summarized as auxiliary functions \cite{heinzel2011auxiliary}).
These auxiliary functions are realized via low-depth phase modulations
of the laser beam using an electro-optic modulator (EOM) \cite{heinzel2011auxiliary}.
Signals for clock synchronization are modulated as sidebands onto
the main carrier. The frequency of the sidebands is chosen such that
interference of the sidebands from the remote and local laser leads
to a sideband/sideband beat note one megahertz away from the main
beat note. This signal is processed separately but not further discussed
here since it is not the focus of the ongoing discussion.\\
Absolute ranging and data transfer are employed using the principle
of direct sequence spread spectrum (DSSS) \cite{sutton2010laser, Esteban:11}.
PRN code sequences are modulated onto the carrier. In turn, data symbols
are modulated onto these code sequences allowing for data transmission.
Since most of the spectral energy of the code modulation lies outside
the narrow bandwidth of the PLL, the chip sequences are not tracked
by the PLL but appear as error in the signal. Consequently, the error
channel (Q-channel) of the PLL is input to a delay-locked loop (DLL)
\cite{sutton2010laser, Esteban:11}. The DLL correlates the incoming
PRN code sequence with a local replica, which indicates the transmission
time of the incoming chip sequence. Moreover, the sign of the correlation
represents the data bit \cite{Delgado2012}. 
\begin{table}
\centering{}\caption{Code delays $\tau$ of the long-arm interferometer for OB $ab$ with
$a,b\in\{1,2,3\}$ and $a\protect\neq b$. \label{tab:Front-End-Delays-ranging}}
\begin{tabular}{ll}
\hline 
\hline Code delay & Symbol\tabularnewline
\hline 
Code generation & $\tau_{\text{Gen},ab}$\tabularnewline
Reference code generation & $\tau_{\text{RGen},ab}$\tabularnewline
Code modulation & $\tau_{\text{Mod},ab}$\tabularnewline
OB delay at transmission (T) at local (L) S/C & $\tau_{\text{OB,T,L},ab,Z}$\tabularnewline
OB delay at reception (R) at remote (R) S/C & $\tau_{\text{OB,R,R},ab,Z}$\tabularnewline
Ranging delay between S/C $b$ and S/C $a$ & $\tau_{\text{R},ab}$\tabularnewline
Code delay of analog front end & $\tau_{\text{FE},ab,\text{L}}$\tabularnewline
PLL processing delay & $\tau_{\text{PLL},ab,\text{L}}$\tabularnewline
DLL processing delay & $\tau_{\text{DLL},ab,\text{L}}$\tabularnewline
Output filtering delay & $\tau_{\text{FO},ab,\text{L}}$\renewcommand{\tabularnewline}{\\ \hline}\tabularnewline
\hline 
\end{tabular}\renewcommand{\tabularnewline}{\\}
\end{table}
Similar to the interferometric phase measurements, the absolute ranging
measurement is subjected to delays. The individual delays have been
located in Fig. \ref{fig:Identification-of-ranging-bias}, denoted
with symbol $\tau$, and identified in Tab. \ref{tab:Front-End-Delays-ranging}.
As a result, the overall delay at the output filter of the DLL is
given by

\begin{align}
\tau_{12}(t & +\tau_{\text{DLL},12,\text{L}}+\tau_{\text{FO},12,\text{L}})\nonumber \\
 & =\tau_{\text{Gen},21}+\tau_{\text{Mod},21}+\tau_{\text{OB,T,L},21,\text{L}}+\tau_{\text{R},12}\nonumber \\
 & +\tau_{\text{OB,R,R},12,\text{L}}+\tau_{\text{FE},12,\text{L}}+\tau_{\text{PLL},12,\text{L}}-\tau_{\text{RGen},12}.\label{eq:tau_0}
\end{align}
Hereby, $\tau_{12}$ denotes the delay measured at the DLL based on
the correlation of the incoming chip sequence with a local replica.
Therefore, delays occurring after the measurement ($\tau_{\text{FO},12,\text{L}}$
and $\tau_{\text{DLL},12,\text{L}}$) appear as delays in the argument
of $\tau_{12}$. The time argument of the code delays has been omitted
in Eq. \ref{eq:tau_0} for readability. \\
Since both, ranging and phase measurements, enter INReP, these
delays must be precisely determined and adequately accounted for in
the data processing of LISA. \\
Delays resulting from the OB ($\delta_{\text{OB}}$ and $\tau_{\text{OB}}$),
are defined through the easily measurable geometric optical path-length
differences (OPDs) on the OB. The latter is made from Zerodur and
exhibits ultra-stable behavior so that any changes in these OPDs are
completely negligible. Similarly, digital delays ($\delta_{\text{PLL},12,Z},\delta_{\text{FO},12,Z}$
with $Z\in\{\text{L},\text{T},\text{R}\}$ and $\tau_{\text{Gen},21},\tau_{\text{RGen},12},\tau_{\text{PLL},12,\text{L}},\tau_{\text{DLL},12,\text{L}},\tau_{\text{FO},12,\text{L}}$)
can be calibrated via on-ground simulations. This is particularly
important for the delay acting on the PRN sequences resulting from
the sequential PLL-DLL architecture \cite{EuringerOptMetr2023,sutton2010laser}.
Finally, the deviation of the test mass from its nominal position
is expected to be on the order of $\approx10^{-4}$ m, resulting
in a sub-picosecond delay. This delay, denoted as $\delta_{\text{TM},12}$
in Eq. \ref{eq:phi_T_0}, is significantly smaller in magnitude compared
to the delays considered hereafter and omitted in the following.\\
At this point, we emphasize that clock differences among the S/C enter
the phase measurement via the ADC conversion and the ranging measurement
in the generation of the PRN code and the reference code. These differences
are addressed separately, in particular via the clock jitter transfer
based on sideband/sideband beat notes, and have been extensively studied
in \cite{reinhardt2023ranging}. Moreover, it has been shown that
these delays can be addressed separately in a stand-alone post-processing
step \cite{Hartwig2022}. Consequently, these delays are assumed to
be known and adequately suppressed.\\
In the following, we will comprise the aforementioned delays, namely
delays from the OB, digital delays, and delays attributed to the clock
differences, as ``known delays'' separating them from the front-end
and modulation delays.\\
Despite the high stability of the phasemeter front end, small variations
are expected to occur due to (1) the variation of the heterodyne frequency
within the large receiver bandwidth (6 to 24 MHz) and due to (2) the
seasonal changes in the environmental temperatures, as well as the
expected ground-to-orbit (GTO) shift. As discussed above, the heterodyne
frequency changes as a consequence of the slowly varying Doppler-shift
and occasional reconfiguration of the offset-locking frequency and
the temperature changes due to variation of the orbital parameters.
Taking these effects into account and based on a simplified model
of the stable front end shown in Fig. \ref{fig:Identification-of-ranging-bias},
we expect the mean front-end delays to be around 20 ns with a variation
on the order of $\pm$5 ns over long timescales of several months.
In addition, ranging measurements are affected by the modulation delay,
which is exposed to similar drifts. On-ground calibration of these
analog delays is complex and time-consuming. Furthermore, on-ground
calibrations are prone to potential errors and likely to change due
to GTO and temperature changes in orbit. Consequently, we seek a possibility
to accurately calibrate and re-calibrate these delays.

\section{Calibration Measurement\label{subsec:Calibration-Measurement}}

In establishing a calibration scheme we first note that ranging
measurements of the long-arm interferometer rely on the principle
of PRN codes modulated onto the carrier with a small modulation depth
\cite{Delgado2012}. As a result, the (main) power of the PRN codes
is in close vicinity to the carrier power spectrum. In addition,
carrier and code signal of the long-arm interferometer are processed
by the identical analog front-end electronics until digitization.
Assuming a sufficiently flat transfer function of the front-end electronic
around the carrier \cite{Barranco2018}, the front-end delay of the
phase $\delta_{\text{FE},12,\text{L}}$ of the long-arm interferometer
and the code $\tau_{\text{FE},12,\text{L}}$ are comparable, i.e.
$\delta_{\text{FE},12,\text{L}}\approx\tau_{\text{FE},12,\text{L}}$.
In fact, this relation holds exactly for phase components up to quadratic
order around the carrier, since the code spectrum is symmetrically
spread around the carrier. Based on the relation, additional ranging
measurements can be introduced to determine the modulation delays
and the front-end delays of all interferometers. These additional
measurements require small modifications to the current baseline,
which will be addressed in the subsequent section \ref{sec:Calibration-performance}.

As indicated in Fig. \ref{fig:Identification-of-interferometer},
the PRN modulation of each laser is performed before the laser output
is split into one part used for local interferometry and another part
that is transmitted to the remote S/C \cite{EESA}. Hence, the beat
notes of the interferometers are modulated by the PRN codes of both,
the laser associated with the respective OB, which we refer to as
``local laser'', as well as the PRN code from another OB, which
we refer to as ``remote laser''. The ``remote laser'' originates
either from the adjacent OB of the same S/C for test-mass and reference
interferometer, or from the OB of the remote S/C for the long-arm
interferometer. Therefore, instead of correlating the incoming signal
of the long-arm interferometer with the code of the distant S/C, a
correlation with the local S/C can be performed. This measurement
gives insight into the transmission path of the local PRN code \cite{Sutton_2013}.
Neglecting the known delays including the digital delays and delays
of the OB, which can simply be subtracted, the measurement reads
\begin{equation}
\text{CL}_{12,L}=\tau_{\text{FE},12,\text{L}}+\tau_{\text{Mod},12}.\label{eq:add_meas_L_CL}
\end{equation}
Hereby, $\text{CL}$, short for ``Calibration Local'', indicates
a calibration (``C'') measurement, where the DLL correlates against
the PRN code modulated onto the laser of the local (``L'') OB. In
contrast, CR, short for ``Calibration Remote'', indicates that the
code modulated onto the laser of the remote (``R'') OB is considered.
In other words, for performing the ``CL'' or ``CR'' measurement,
the local or the remote code must be used by the reference code generator
of the DLL, respectively. Finally, the subscript indicates the OB
and the interferometer, separated by a comma.

In the same way, the incoming signal of the reference and test-mass
interferometer features a PRN modulation and the error channel of
the PLL reveals these chip sequences. Forwarding this signal to a
DLL enables us to extend the application of ranging measurements to
gain insight into the delays present in the read-out of the test-mass
and reference interferometer. Following this approach, we introduce
the additional calibration measurements
\begin{align}
\text{CR}_{12,T} & =\tau_{\text{BL},12}+\tau_{\text{FE},12,\text{T}}+\tau_{\text{Mod},13},\label{eq:add_meas_T_CR}\\
\text{CR}_{12,R} & =\tau_{\text{BL},12}+\tau_{\text{FE},12,\text{R}}+\tau_{\text{Mod},13},\label{eq:add_meas_R_CR}\\
\text{CL}_{12,R} & =\tau_{\text{FE},12,\text{R}}+\tau_{\text{Mod},12},\label{eq:add_meas_R_CL}
\end{align}

where similarly to Eq. \ref{eq:add_meas_L_CL}, known delays have
been omitted. The very same measurements can be performed on OB 13.
Importantly, the fiber backlink connecting both OBs is bi-directional
and requires reciprocal phase stability \cite{Fleddermann_2018},
which will be accounted for as $\tau_{\text{BL},12}=\tau_{\text{BL},13}$
in the following. As a result, the system of equations consists of
8 calibration measurement equations with 9 delays as variables. Although
the system of equations is under-determined, it can be solved by expressing
the delays in terms of one common delay. This delay can be arbitrarily
chosen and shall be referred to as ``reference delay'' in the following.
Exemplarily, Eq. \ref{eq:calibration_matrix_xy-1} expresses the
delays in terms of the front-end delay of the long-arm interferometer
of OB 12, $\tau_{\text{FE},12,L}$, which serves as the ``reference
delay'' of S/C 1. \begin{widetext} \small

\begin{equation}
\left(\begin{array}{c}
\tau_{\text{FE},12,\text{T}}\\
\tau_{\text{FE},12,\text{R}}\\
\tau_{\text{FE},13,\text{L}}\\
\tau_{\text{FE},13,\text{T}}\\
\tau_{\text{FE},13,\text{R}}\\
\tau_{\text{Mod},12}\\
\tau_{\text{Mod},13}\\
\tau_{\text{BL},12}
\end{array}\right)=\left(\begin{array}{c}
-\text{CL}_{12,\text{L}}+\text{CL}_{12,\text{R}}-\text{CR}_{12,\text{R}}+\text{CR}_{12,\text{T}}+\tau_{\text{FE},12,\text{L}}\\
-\text{CL}_{12,\text{L}}+\text{CL}_{12,\text{R}}+\tau_{\text{FE},12,\text{L}}\\
\frac{1}{2}\left(-2\text{CL}_{12,\text{L}}+2\text{CL}_{13,\text{L}}+\text{CL}_{12,\text{R}}-\text{CL}_{13,\text{R}}-\text{CR}_{12,\text{R}}+\text{CR}_{13,\text{R}}\right)+\tau_{\text{FE},12,\text{L}}\\
\frac{1}{2}\left(-2\text{CL}_{12,\text{L}}+\text{CL}_{12,\text{R}}+\text{CL}_{13,\text{R}}-\text{CR}_{12,\text{R}}-\text{CR}_{13,\text{R}}+2\text{CR}_{13,\text{T}}\right)+\tau_{\text{FE},12,\text{L}}\\
\frac{1}{2}\left(-2\text{CL}_{12,\text{L}}+\text{CL}_{12,\text{R}}+\text{CL}_{13,\text{R}}-\text{CR}_{12,\text{R}}+\text{CR}_{13,\text{R}}\right)+\tau_{\text{FE},12,\text{L}}\\
\text{CL}_{12,\text{L}}-\tau_{\text{FE},12,\text{L}}\\
\frac{1}{2}\left(2\text{CL}_{12,\text{L}}-\text{CL}_{12,\text{R}}+\text{CL}_{13,\text{R}}+\text{CR}_{12,\text{R}}-\text{CR}_{13,\text{R}}\right)-\tau_{\text{FE},12,\text{L}}\\
\frac{1}{2}\left(-\text{CL}_{12,\text{R}}-\text{CL}_{13,\text{R}}+\text{CR}_{12,\text{R}}+\text{CR}_{13,\text{R}}\right)
\end{array}\right)=\left(\begin{array}{c}
m_{\text{FE},12,\text{T}}+\tau_{\text{FE},12,\text{L}}\\
m_{\text{FE},12,\text{R}}+\tau_{\text{FE},12,\text{L}}\\
m_{\text{FE},13,\text{L}}+\tau_{\text{FE},12,\text{L}}\\
m_{\text{FE},13,\text{T}}+\tau_{\text{FE},12,\text{L}}\\
m_{\text{FE},13,\text{R}}+\tau_{\text{FE},12,\text{L}}\\
m_{\text{Mod},12}-\tau_{\text{FE},12,\text{L}}\\
m_{\text{Mod},13}-\tau_{\text{FE},12,\text{L}}\\
m_{\text{BL},12}
\end{array}\right)\label{eq:calibration_matrix_xy-1}
\end{equation}
\end{widetext} \normalsize
In the expressions on the right surrounded by round brackets, measurement
parameters $m$ with respective subscripts are introduced, which represent
a combination of the individual calibration parameters. We note that
all front-end delays follow the same scheme where the reference delay
is added to all measurement parameters $m$. In contrast, the modulation
delays consist of the difference between the measurement parameters
$m$ and the reference delay. Finally, the fiber backlink delay is
directly retrieved via the measurement parameter. Therefore, this
delay is considered as ``known'' and omitted in the following discussion.
Importantly, the calibration scheme allows expressing the modulation
and front-end delays of all interferometers in terms of an individual
measurement parameter $m$ and one common reference delay per S/C.

\section{TDI Processing\label{sec:TDI-Processing}}

TDI is a data post-processing technique designed to mitigate the
impact of laser frequency noise by precisely time-shifting and linearly
combining the interferometer measurements from multiple S/C. It is
part of INReP, a crucial component within the LISA data analysis framework,
responsible for the initial processing and filtering of the acquired
data. INReP prepares the measurements for gravitational wave analysis.
To assess the impact of the calibration scheme outlined in the preceding
section on TDI, we provide a short summary of INReP and the TDI algorithm
with reference to \cite{OttoThesis2015}. For a more comprehensive
understanding of INReP and the TDI algorithm, the reader is referred
to \cite{Hartwig2022,HartwigThesis2021,HoubaJoG}.\\
In section \ref{subsec:Ranging-Biases}, the delays present in the
phase (Eqs. \ref{eq:phi_L_0} to \ref{eq:phi_R_0}) and ranging (Eq.
\ref{eq:tau_0}) measurements have been identified. In the following,
we focus explicitly on the unknown analog delays, namely the front-end
and the modulation delays, that have been addressed by the calibration
scheme introduced in section \ref{subsec:Calibration-Measurement}.
The known delays in the ranging measurements can simply be subtracted
from Eq. \ref{eq:tau_0} and thus do not further affect the measurements.
This is also the case for delays in the interferometric measurements
(Eqs. \ref{eq:phi_L_0} to \ref{eq:phi_R_0}), that appear equally
in the beams of the interfering lasers. In fact, delays that occur
exclusively in one laser must be examined separately. While this is
the subject of ongoing investigations, it will not be further explored
in the ongoing analysis.\\
Following this approach, Eqs. \ref{eq:phi12L_p} to \ref{eq:phi12R_p}
illustrate the laser noise present in the measurements of the long-arm,
test-mass, and reference interferometers introduced by Eqs. \ref{eq:phi_L_0}
to \ref{eq:phi_R_0}.

\begin{align}
\Phi_{12,L}^{\text{p}}(t) & =p_{21}(t-\tau_{\text{R},12}-\tau_{\text{FE},12,\text{L}})-p_{12}(t-\tau_{\text{FE},12,\text{L}}),\label{eq:phi12L_p}\\
\Phi_{12,T}^{\text{p}}(t) & =p_{13}(t-\tau_{\text{FE},12,\text{T}})-p_{12}(t-\tau_{\text{FE},12,\text{T}}),\label{eq:phi12T_p}\\
\Phi_{12,R}^{\text{p}}(t) & =p_{13}(t-\tau_{\text{FE},12,\text{R}})-p_{12}(t-\tau_{\text{FE},12,\text{R}}).\label{eq:phi12R_p}
\end{align}
Hereby, the superscript $\text{p}$ in the expressions $\Phi_{12,L}^{\text{p}}(t)$,
$\Phi_{12,T}^{\text{p}}(t)$, and $\Phi_{12,R}^{\text{p}}(t)$, represents
the impact of laser noise $p_{ab}(t)$ on the interferometric phase
measurements. Moreover, phase delays have been expressed via the code
delays, assuming a sufficiently flat transfer function in the vicinity
of the carrier as explained in section \ref{subsec:Calibration-Measurement}.
The influence of laser noise on the remaining five OBs can be obtained
from Eqs. \ref{eq:phi12L_p} to \ref{eq:phi12R_p} through a cyclic
permutation of the indices.

In LISA, the set of interferometric measurements, along with the measured
light travel times between the S/C are processed by INReP. Before
the execution of TDI, additional algorithms are carried out to mitigate
translational OB displacement noise, suppress clock noise, and reduce
the number of laser noise sources from six to three. These algorithms
are elaborated in \cite{OttoThesis2015}. Importantly, all of the
individual algorithms rely on linear combinations of the various phase
measurements and shifting of these phase measurements in time. The
shift in time is incorporated via a time-shift operator. In the following,
we will provide an adapted definition of this time-shift operator,
which allows for a compensation of front-end and modulation delays.
Importantly, the compensation principle is demonstrated based on the
time-shift operator itself, and since all of the individual algorithms
incorporate this operator, compensation also applies to these algorithms.\\
In order to demonstrate the suppression of laser noise as given in
Eqs. \ref{eq:phi12L_p} to \ref{eq:phi12R_p}, it is sufficient to
perform only the algorithm to reduce the number of laser noise sources
from six to three before entering the TDI algorithm. However, the
validity of the results persists when implementing the full INReP
algorithm, as substantiated by arguments presented in the preceding
paragraph. Equations \ref{eq:eta12_p} to \ref{eq:eta31_p} characterize
the algorithm for the reduction of the number of laser noise sources
from six to three required to obtain the TDI second-generation Michelson
$X(t)$ channel.
\begin{align}
\eta_{12}^{\text{p}}(t) & =\Phi_{12,L}^{\text{p}}-D_{12}\left(\dfrac{\Phi_{23,R}^{\text{p}}-\Phi_{21,R}^{\text{p}}}{2}\right)\label{eq:eta12_p}\\
\eta_{13}^{\text{p}}(t) & =\Phi_{13,L}^{\text{p}}-\dfrac{\Phi_{12,R}^{\text{p}}-\Phi_{13,R}^{\text{p}}}{2}\label{eq:eta13_p}\\
\eta_{21}^{\text{p}}(t) & =\Phi_{21,L}^{\text{p}}-\dfrac{\Phi_{23,R}^{\text{p}}-\Phi_{21,R}^{\text{p}}}{2}\label{eq:eta21_p}\\
\eta_{31}^{\text{p}}(t) & =\Phi_{31,L}^{\text{p}}-D_{31}\left(\dfrac{\Phi_{12,R}^{\text{p}}-\Phi_{13,R}^{\text{p}}}{2}\right)\label{eq:eta31_p}
\end{align}

Hereby, $D_{ba}$ represents the time shift operator that shifts a
measurement performed on S/C $a$ by the time it takes for a signal
to be received on S/C $b$. \\
TDI encompasses various combinations, such as Michelson, Sagnac, and
further derivatives, offering distinct advantages in gravitational
wave analysis \cite{Tinto_2004}. The TDI second-generation Michelson
variable considered in this paper is a four-link combination requiring
measurements from OB 12, OB 13, OB 21, and OB 31. The underlying algorithm
is 
\begin{align}
X(t) & =\eta_{13}^{\text{p}}-\eta_{12}^{\text{p}}+D_{A}\eta_{31}^{\text{p}}-D_{T}\eta_{21}^{\text{p}}\nonumber \\
 & +D_{B}\eta_{12}^{\text{p}}-D_{U}\eta_{13}^{\text{p}}+D_{C}\eta_{21}^{\text{p}}-D_{V}\eta_{31}^{\text{p}}\nonumber \\
 & +D_{D}\eta_{12}^{\text{p}}-D_{W}\eta_{13}^{\text{p}}+D_{E}\eta_{21}^{\text{p}}-D_{X}\eta_{31}^{\text{p}}\nonumber \\
 & +D_{F}\eta_{13}^{\text{p}}-D_{Y}\eta_{12}^{\text{p}}+D_{G}\eta_{31}^{\text{p}}-D_{Z}\eta_{21}^{\text{p}}.\label{eq:TDI20X}
\end{align}
The time argument has been omitted above for readability. Note that
multiple, nested delays are necessary to effectively cancel laser
noise. Thereby, the nested delay operator is defined as $D_{ab,cb}f(t)=D_{cb}D_{ab}f(t)$
and similar for $n$-fold delay operations. The delay arrangements
have been summarized as letters for clarity. The definitions of the
subscript delay letters can be found in Table \ref{tab:translationtable}
\cite{houba2022lisa}.
\begin{table}
\centering{}\caption{Translation Table for the Delay Index Notation.\label{tab:translationtable}}
\begin{tabular}{cccc}
\hline 
\hline Index & Delays & Index & Delays\tabularnewline
\hline 
$A$ & 13 & $W$ & 31,13,21,12\tabularnewline
$T$ & 12 & $E$ & 12,21,12,31,13\tabularnewline
$B$ & 31,13 & $X$ & 13,31,13,21,12\tabularnewline
$U$ & 21,12 & $F$ & 21,12,21,12,31,13\tabularnewline
$C$ & 12,31,13 & $Y$ & 31,13,31,13,21,12\tabularnewline
$V$ & 13,21,12 & $G$ & 13,21,12,21,12,31,13\tabularnewline
$D$ & 21,12,31,13 & $Z$ & 12,31,13,31,13,21,12\renewcommand{\tabularnewline}{\\ \hline}\tabularnewline
\hline 
\end{tabular}\renewcommand{\tabularnewline}{\\}
\end{table}

The current TDI processing considers the S/C as a point mass, which
implies that all interferometric measurements per S/C are performed
at the same location. The time-shift operator is thereby defined according
to \cite{HartwigThesis2021}
\begin{equation}
D_{ba}^{\text{ideal}}f(t):=f\left(t-\tau_{\text{R},ba}\right),\label{eq:D_ideal_ba}
\end{equation}
to evaluate an arbitrary measurement $f(t)$ of S/C $a$ at the time
of reception on S/C $b$ with $a,b\in\{1,2,3\}$ and $a\neq b$. We
refer to these generic indices since INReP processing takes into account
measurements from all 3 S/C. Indeed, definition \ref{eq:D_ideal_ba}
yields a static time-shift operator, as it excludes the consideration
of the time dependence in the ranging delay $\tau_{\text{R},ba}$.
Nevertheless, we will demonstrate that the outcomes remain valid even
when accounting for time-dependent ranging delays.\\
Definition \ref{eq:D_ideal_ba} faces two problems when taking into
account the analog delays in the phase and ranging measurements:

(1) The interferometric measurements and thus the laser noise are
delayed by individual front-end delays, cf. Eqs. \ref{eq:phi12L_p}
to \ref{eq:phi12R_p}, which makes the consideration of S/C as a point
mass incorrect. This circumstance can be addressed by incorporating
the calibration scheme of Eq. \ref{eq:calibration_matrix_xy-1}, enabling
the expression of individual front-end delays through a reference
delay and an individual measurement parameter $m$. In the following,
we shift all phase measurements by this individual measurement delay.
Exemplary, for the laser noise of the test-mass interferometer of
OB 12, cf. Eq. \ref{eq:phi12T_p}, this results in
\begin{align}
\Phi_{12,T}^{\text{p}} & (t+m_{\text{FE},12,T})\nonumber \\
 & =p_{13}(t-\tau_{\text{FE},12,\text{T}}+m_{\text{FE},12,T})\nonumber \\
 & -p_{12}(t-\tau_{\text{FE},12,\text{T}}+m_{\text{FE},12,T})\nonumber \\
 & =p_{13}(t-(m_{\text{FE},12,\text{T}}+\tau_{\text{FE},12,\text{L}})+m_{\text{FE},12,T})\nonumber \\
 & -p_{12}(t-(m_{\text{FE},12,\text{T}}+\tau_{\text{FE},12,\text{L}})+m_{\text{FE},12,T})\nonumber \\
 & =p_{13}(t-\tau_{\text{FE},12,\text{L}})-p_{12}(t-\tau_{\text{FE},12,\text{L}}),\label{eq:Phi_shift_T_12}
\end{align}
where we used the result of Eq. \ref{eq:calibration_matrix_xy-1},
row 1. Consequently, the interferometric measurements per S/C are
delayed by a single reference delay and definition \ref{eq:D_ideal_ba}
would still be accurate.

(2) Knowledge about the distance $\tau_{\text{R}}$ among the S/C
is obtained via the absolute ranging measurement $\tau_{ba}$ of the
DLL, cf. Eq. \ref{eq:tau_0}, which is affected by additional delays.
Omitting the known delays, the ranging measurement is given by
\begin{equation}
\tau_{ba}=\tau_{\text{Mod},ab}+\tau_{\text{R},ba}+\tau_{\text{FE},ba,\text{L}}.\label{eq:t_ba_mod}
\end{equation}
Similar to the phase measurement, we can express the additional
delays, i.e. the modulation and front-end delay, via a reference
delay taking into account row 6 and row 3 of Eq. \ref{eq:calibration_matrix_xy-1},
respectively. This results in
\begin{align}
\tau_{\text{Mod},ab} & =m_{\text{Mod},ab}-\tau_{\text{FE},ac,\text{L}},\label{eq:tau_mod_ab}\\
\tau_{\text{FE},ba,\text{L}} & =m_{\text{FE},ba,\text{L}}+\tau_{\text{FE,}bd,\text{L}},\label{eq:tau_fe_ba}
\end{align}
with $a,b,c,d\in\{1,2,3\}$ and $b\neq a\neq c,b\neq d$. Without
loss of generality, we considered the front-end delays $\tau_{\text{FE},ac,\text{L}}$
and $\tau_{\text{FE,}bd,\text{L}}$ of the long-arm interferometers
as the reference delays of S/C $a$ and $b$, respectively. Following
this procedure, the measurement of Eq. \ref{eq:t_ba_mod} is shifted
by the measurement parameters $m_{\text{Mod},ab}$ and $m_{\text{FE},ba,\text{L}}$
of Eq. \ref{eq:tau_mod_ab} and Eq. \ref{eq:tau_fe_ba}, respectively.
Therefore, we account for these known measurement parameters by defining
a biased time-shift operator as 
\begin{equation}
D_{ba}^{\text{bias}}f(t):=f(t-\tau_{ba}+m_{\text{Mod},ab}+m_{\text{FE},ba,\text{L}}).\label{eq:D_biased_ba}
\end{equation}

At this point, we note that the time-shift operator $D_{ba}$ is applied
only to an interferometric measurement performed on S/C $a$ to evaluate
it at the time of reception onboard S/C $b$. Based on the mitigation
of problem (1), all phase measurements of S/C $a$ are expressed in
terms of one reference delay and application of the the biased time-shift
operator $D_{ba}^{\text{bias}}$ yields\\
\begin{align}
D_{ba}^{\text{bias}}f(t & -\tau_{\text{FE},ac,\text{L}})\nonumber \\
=f(t & -\tau_{\text{FE},ac,\text{L}}-\tau_{ba}+m_{\text{Mod,ab}}+m_{\text{FE},ba,\text{L}})\nonumber \\
=f(t & -\tau_{\text{FE},ac,\text{L}}-\tau_{\text{Mod},ab}-\tau_{\text{R},ba}\nonumber \\
 & -\tau_{\text{FE},ba,\text{L}}+m_{\text{Mod},ab}+m_{\text{FE},ba,\text{L}})\nonumber \\
=f(t & -\tau_{\text{FE},ac,\text{L}}-m_{\text{Mod},ab}+\tau_{\text{FE},ac,\text{L}}-\tau_{\text{R},ba}\nonumber \\
 & -m_{\text{FE},ba,\text{L}}-\tau_{\text{FE,}bd,\text{L}}+m_{\text{Mod},ab}+m_{\text{FE},ba,\text{L}})\nonumber \\
=f(t & -\tau_{\text{R},ba}-\tau_{\text{FE,}bd,\text{L}}).\label{eq:d_bias_deriv}
\end{align}
We observe, the biased time-shift operator applies the correct ranging
delay but also changes the front-end delay to the S/C, where the measurement
shall be evaluated. In fact, we can express the biased time-shift
operator in terms of the ideal time-shift operator by combining Eq.
\ref{eq:D_ideal_ba} and Eq. \ref{eq:d_bias_deriv},
\begin{equation}
D_{ba}^{\text{bias}}f(t-\tau_{\text{FE},ac,\text{L}})=D_{ba}^{\text{ideal}}f(t-\tau_{\text{FE,}bd,\text{L}}).\label{eq:D_bias-D_ideal}
\end{equation}
Importantly, nesting the time-shift operator results in
\begin{equation}
D_{cb}^{\text{bias}}D_{ba}^{\text{bias}}f(t-\tau_{\text{FE},ac,\text{L}})=D_{cb}^{\text{ideal}}D_{ba}^{\text{ideal}}f(t-\tau_{\text{FE},ce,\text{L}}).\label{eq:Recursive applicaiton biased time shift}
\end{equation}
Again the reference delay of S/C $c$ is arbitrarily chosen to be
$\tau_{\text{FE},ce,\text{L}}$. Finally, the indices in Eqs. \ref{eq:D_biased_ba}
to \ref{eq:Recursive applicaiton biased time shift} are defined by
$a,b,c,d,e\in\{1,2,3\}$ and $a\neq b\neq c,a\neq c\neq e,b\neq d$.
Following this observation, we can conclude that by nested application
of the biased time-shift operator, only the front-end delay of the
S/C, where the measurement shall be evaluated at last, remains. In
the following, we will refer to this last reference delay as the leading
reference delay. \\
So far, the time-shift operator has been applied in a static manner,
meaning that the delays $\tau$ are time-invariant, an assumption
applied in TDI 1.0. In contrast, the dynamic time shift operator used
for TDI 2.0 assumes a time variation in the distances among the S/Cs.
Nesting of the dynamic time-shift operator results in additional reference
delays in the argument of the function $f(t)$. However, as shown
in appendix \ref{sec:Appendix:-Biased-Dynamic}, these additional
contributions are on the order of femtoseconds. Hence, Eq. \ref{eq:Recursive applicaiton biased time shift}
also holds in good approximation for the dynamic time shift operator.

Based on the behavior of nested biased time-shift operators as given
in Eq. \ref{eq:Recursive applicaiton biased time shift}, analysis
of the biased time-shift operator on INReP can be readily performed.
Closely inspecting the terms that constitute the TDI $X$ variable
in Eq. \ref{eq:TDI20X}, reveals that the leading reference delay
is identical for every term, namely $\tau_{\text{FE},12,\text{L}}$.
\begin{figure}
\centering{}\includegraphics[width=1\linewidth]{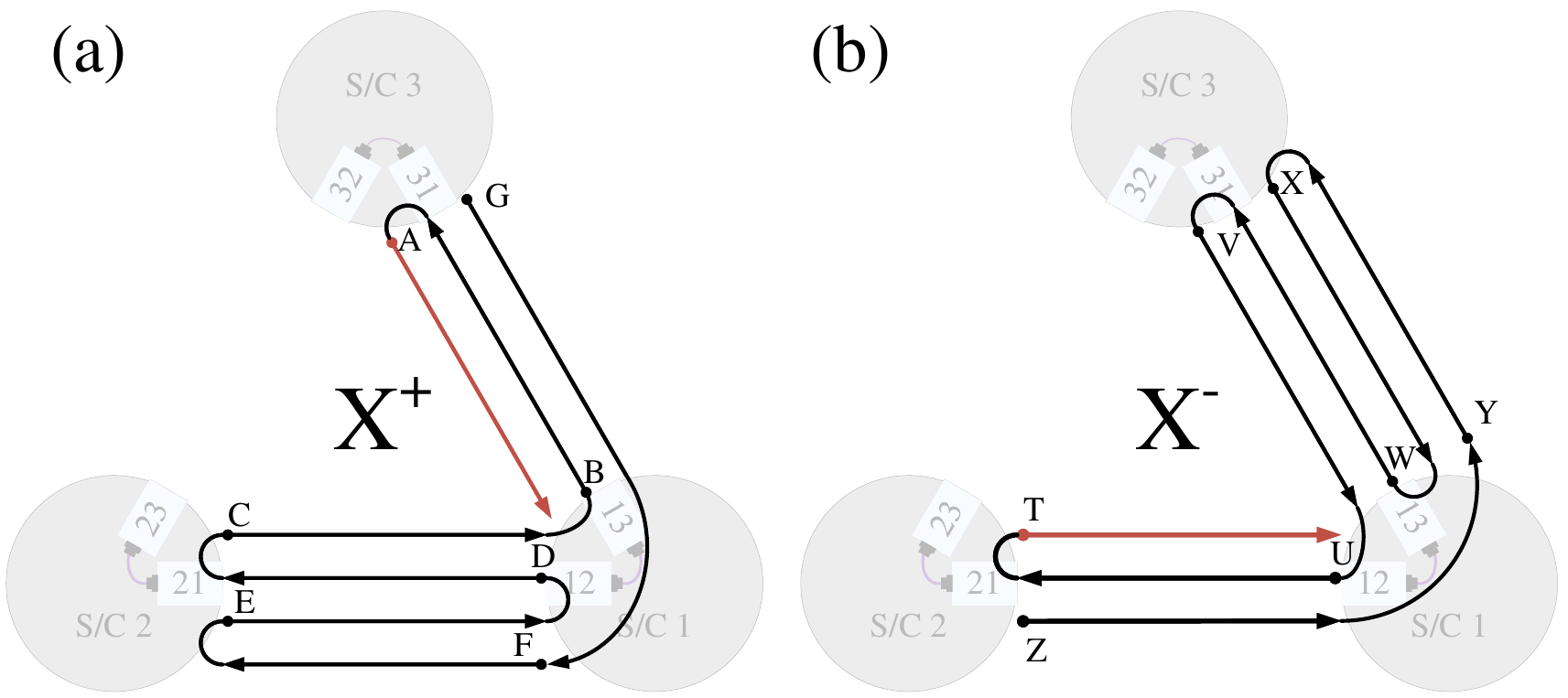}\caption{Schematic indicating the routing of the (nested) delay operators applied
to the individual $\eta$-terms composing the TDI $X$ variable. The
letters refer to the subscripts of the delay operators as given in
Eq. \ref{eq:TDI20X} and indicate the start of the respective (nested)
delay operator. The $\eta$-terms entering with a positive and negative
sign have been separated in panels (a) and (b), respectively.\label{fig:TDI-Michelson-Variables.}}
\end{figure}
This observation can also be obtained graphically by illustrating
the routing of the delay operators applied to the individual $\eta$-terms
composing the TDI $X$ variable in Eq. \ref{eq:TDI20X}. Hereby, terms
entering with a positive and negative sign have been separated in
Fig. \ref{fig:TDI-Michelson-Variables.}(a) and (b), respectively.
The subscript of the (nested) delay operators, as given in Tab. \ref{tab:translationtable},
indicates the start of the respective (nested) delay operator, while
arrows represent the individual delay operators. Focusing on Fig.
\ref{fig:TDI-Michelson-Variables.}(a), we note that the paths created
by the arrows are continuous and the (nested) delay operators share
the same route. As a result, all (nested) delay operators end at the
same S/C, namely at S/C 1. The same observation applies to Fig. \ref{fig:TDI-Michelson-Variables.}(b).
We can conclude that all terms entering the composition of the TDI
$X$ variable are shifted by one common delay, the reference delay
of S/C 1 $\tau_{\text{FE},12,\text{L}}$. Since, the remaining TDI
Michelson variables $Y$ and $Z$ can be obtained from a cyclic permutation
of measurement and delay indices one finds that these variables are
delayed by the reference delay of S/C 2 and S/C 3, respectively. In
this sense, the application of the calibration scheme translates the
individual delays of the interferometers to a common delay per TDI
Michelson variable and therefore -- exact calibration provided --
the individual delays do not affect the measurement performance. \\
Moreover, this principle also applies to higher-order Michelson variables,
where the distinction from lower-order variables depends on the number
of loops considered by the nested delay operators, while the leading
reference delay remains unchanged \cite{tinto2023higherorder}.

\section{Calibration performance\label{sec:Calibration-performance}}

Utilizing the calibration method, the estimation of the delays relies
on code measurements. The accuracy of these measurements depends
on the code tracking configuration, in particular on the bandwidth
and the discriminator of the DLL, and the signal-to-noise ratio.\\
The current baseline of LISA does not take into account calibration
measurements. 
\begin{figure}
\centering{}\includegraphics[width=1\linewidth]{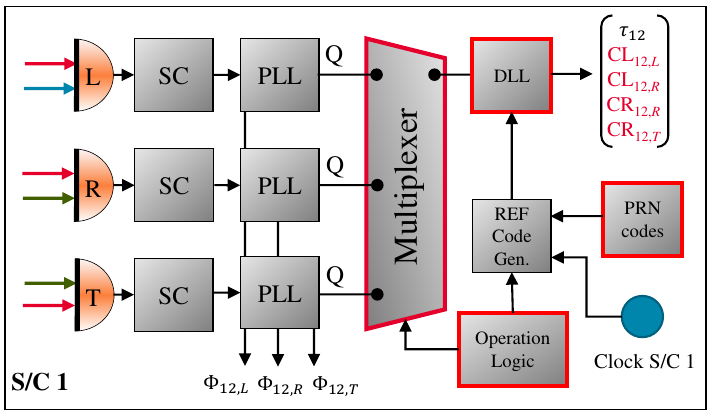}\caption{Schematic indicating via red frames the changes to the current baseline,
necessary for the implementation of the proposed calibration scheme.
The calibrations are controlled by an operation logic, which specifies
the PRN code used in the reference code generator (``REF Code Gen.'')
and the error channel (``Q'') that is routed to the DLL via a multiplexer.
The processing electronic, TIA and ADC, is summarized as signal conditioning
(``SC''). \label{fig:Multiplexer-Scheme.}}
\end{figure}
Therefore, the least invasive method to perform these measurements
is by means of time multiplexing, where all calibration measurements
are performed using the DLL of the long-arm interferometer at fixed
intervals. This avoids the need for additional DLLs but excludes
the possibility of performing continuous calibrations. As outlined
in section \ref{subsec:Ranging-Biases}, front-end delay variations
are primarily attributed to temperature drifts, based on seasonal
changes, and spectral drifts which are slowly varying on the timescale
of weeks or months \cite{EESA}. Consequently, intervals on the time
scale of weeks seem to be sufficient. During the fixed calibration
session the calibration of the long-arm, test-mass and reference interferometers
of the S/C shall be performed sequentially. This process involves
the transfer of the PLL error channel, denoted as ``Q'' in Fig.
\ref{fig:Multiplexer-Scheme.}, from the interferometer used for calibration
to the DLL situated at the phasemeter of the long-arm interferometer
via a multiplexer.\\
In order to increase the calibration performance, only the laser whose
transmission path is currently subject to calibration shall be code-modulated,
i.e. the PRN code of the other interfering laser is constantly set
to one. Moreover, the code used for the calibration shall be free
of any data modulation, enabling a coherent early-late correlation
in the DLL among the whole chip sequence. Finally, the calibration
time per measurement is assumed to be 100 seconds.
\begin{table}
\centering{}\caption{Baseline parameters\label{tab:Baseline-Parameter-1}}
\begin{tabular}{lll}
\hline 
\hline Parameter (Unit) & Symbol & Value\tabularnewline
\hline 
Modulation index (-) & $m_{\text{prn}}$ & 0.1\tabularnewline
Chip period (ms) & $T_{c}$ & 0.001\tabularnewline
Sampling rate (MHz) & $f_{s}$ & 80\tabularnewline
Symbol period (ms) & $T_{s}$ & 1.024\tabularnewline
Ranging bandwidth (Hz) & $B_{\text{DLL}}$ & 4\tabularnewline
BOC(1,1) early-late spacing ($T_{c}$) & $\Delta_{\text{BOC}}$ & 0.2\renewcommand{\tabularnewline}{\\ \hline}\tabularnewline
\hline 
\end{tabular}\renewcommand{\tabularnewline}{\\}
\end{table}

It is worth noting that the absence of data modulation leads to the
fact, that the sequential PLL-DLL architecture does not induce ranging
variations. Instead, only a fixed ranging delay in the order of nanoseconds
is present, which can be calibrated via simulations on ground and
thus does not contribute to the calibration error \cite{EuringerTIM2023}.
The performance of the calibration measurements is thus limited by
two effects. (1) The incoming PRN sequences are contaminated with
noise. The dominant noise source strongly depends on the interferometer.
While the long-arm interferometer is limited by shot noise, this noise
source plays a subordinate role in the test mass and reference interferometer,
where stray light is one major noise source \cite{LISAPRFMODEL}.
The stray light contribution increases at low frequencies, however,
this increase is counteracted by the high-pass behavior of the transfer
function in the PLL error channel, which serves as input for the DLL
\cite{Delgado2012}. Consequently, these noise contributions have
been modeled as white noise. The error resulting from these contributions
is denoted as $\sigma_{\text{n,c}}$ and can be estimated based on
the analytical formulas of Betz \cite{BetzGen2009,BetzII2009}. Hereby,
the subscript ``c'' denotes the coherent processing inside the DLL.
(2) In addition, the DLL is working in the digital domain, causing
performance degradation due to the granularity of the PRN code generator.
This error is represented via $\sigma_{\text{o},\pm}$, where ($+$)
denotes a maximum and ($-$) a minimum error and was derived in \cite{EuringerOptMetr2023}.
Since both effects are uncorrelated the combined rms (root-mean-square)
calibration error is given by
\begin{align}
\sigma_{\text{c}} & \approx\sqrt{\sigma_{\text{o},+}^{2}+\sigma_{\text{n,c}}^{2}},\label{eq:rms_calib_error}
\end{align}
where the application of $\sigma_{\text{o},+}$ leads to a conservative
approach. Based on these assumptions and the parameters specified
in Tab. \ref{tab:Baseline-Parameter-1}, the calibration performance
can be estimated. Thereby, LISA representative noise values from \cite{LISAPRFMODEL}
have been taken into account for each interferometer. The resulting
rms calibration errors are in the order of millimeters and dominated
by the noise contributions $\sigma_{\text{n,c}}$ of the incoming
signal. Individual calibration errors have been listed in Tab. \ref{tab:RMS-calibration-error}.
\begin{table}
\begin{centering}
\caption{RMS errors\label{tab:RMS-calibration-error}}
\begin{tabular}{cc}
\hline 
\hline Measurement & RMS error\tabularnewline
\hline 
Calibration of long-arm interferometer & 3.3 mm\tabularnewline
Calibration of test-mass interferometer & 1.3 mm\tabularnewline
Calibration of reference interferometer & 0.7 mm\tabularnewline
Ranging & 12 cm\tabularnewline
Ranging when wiping local code & 6.6 cm\renewcommand{\tabularnewline}{\\ \hline}\tabularnewline
\hline 
\end{tabular}\renewcommand{\tabularnewline}{\\}
\par\end{centering}
\end{table}
\\
In a similar manner, the ranging error of the long-arm interferometer
can be estimated. However, two additional effects need to be considered.
First of all, the interfering code of the local laser introduces an
additional error $\sigma_{\text{i}}$, which has also been addressed
by Betz \cite{BetzGen2009,BetzII2009}. Secondly, the sequential PLL-DLL
processing in combination with data modulation of the chip sequences
leads to ranging delay variations in the DLL \cite{EuringerTIM2023}.
However, for the sampling period considered (80 MHz, cf. Tab. \ref{tab:Baseline-Parameter-1}),
these variations are much smaller than those arising from the granularity
of the PRN code generator \cite{EuringerOptMetr2023}. Thus the ranging
error is given by
\begin{align}
\sigma_{\text{r}} & \approx\sqrt{\sigma_{\text{o},+}^{2}+\sigma_{\text{n,n}}^{2}+\sigma_{\text{i,n}}^{2}}.\label{eq:rms_ranging_error}
\end{align}
Thereby, the additional subscript ``n'' in $\sigma_{\text{n,n}}$
and $\sigma_{\text{i,n}}$ denotes the non-coherent processing of
the DLL, which is necessary to remove the data modulation. Based
on the parameters specified in Tab. \ref{tab:Baseline-Parameter-1},
the rms ranging errors are in the order of several centimeters and
listed in Tab. \ref{tab:RMS-calibration-error}.

\section{Simulation and Results\label{sec:Simulation-and-Results}}

In order to assess the influence of the front-end and modulation delays
and to verify the proposed calibration scheme, numerical simulations
have been performed. The simulation comprises two stages: signal generation
and signal processing. In the first stage, the interferometric measurements
are generated according to Eqs. \ref{eq:phi12L_p} to \ref{eq:phi12R_p}.
Uncorrelated laser noise is considered for each OB as per the requirement.
The light travel times are assumed to be time-varying, with relative
S/C velocities of around 15 m/s \cite{OttoThesis2015}. In the signal
processing stage, all steps of INReP are executed, even though for
the simulations in this paper, the noise sources to be suppressed
by INReP, apart from laser noise, are set to zero. Front-end delays
and modulation delays are taken into account in the generation of
ranging measurements as well as interferometric signals. These delays
are considered random but constant with respect to the simulation
time, as described in section \ref{subsec:Ranging-Biases}. Therefore,
the values of the individual front-end delays of the interferometers
and the modulation delays have been drawn from a uniform distribution
within an interval of 20$\pm$5 ns and 10$\pm$5 ns, respectively.
Moreover, calibration errors $\sigma_{\text{c}}$ and ranging errors
$\sigma_{\text{r}}$ are taken into account as specified in Tab. \ref{tab:RMS-calibration-error}
unless otherwise noted. Signal processing is assessed for three primary
scenarios regarding front-end and modulation delays: (1) uncompensated
delays, (2) compensation of the mean delays, representing a single,
ground-based calibration, and (3) compensation of the delays including
variations, representing the execution of periodic calibration measurements
performed in space. Depending on the simulation scenarios, the calibration
scheme of section \ref{subsec:Calibration-Measurement} is used by
INReP to obtain the front-end and modulation delay estimates and mitigate
their impact on residual laser noise in the TDI second-generation
Michelson combination.
\begin{figure*}
\centering{}\includegraphics[width=0.95\linewidth]{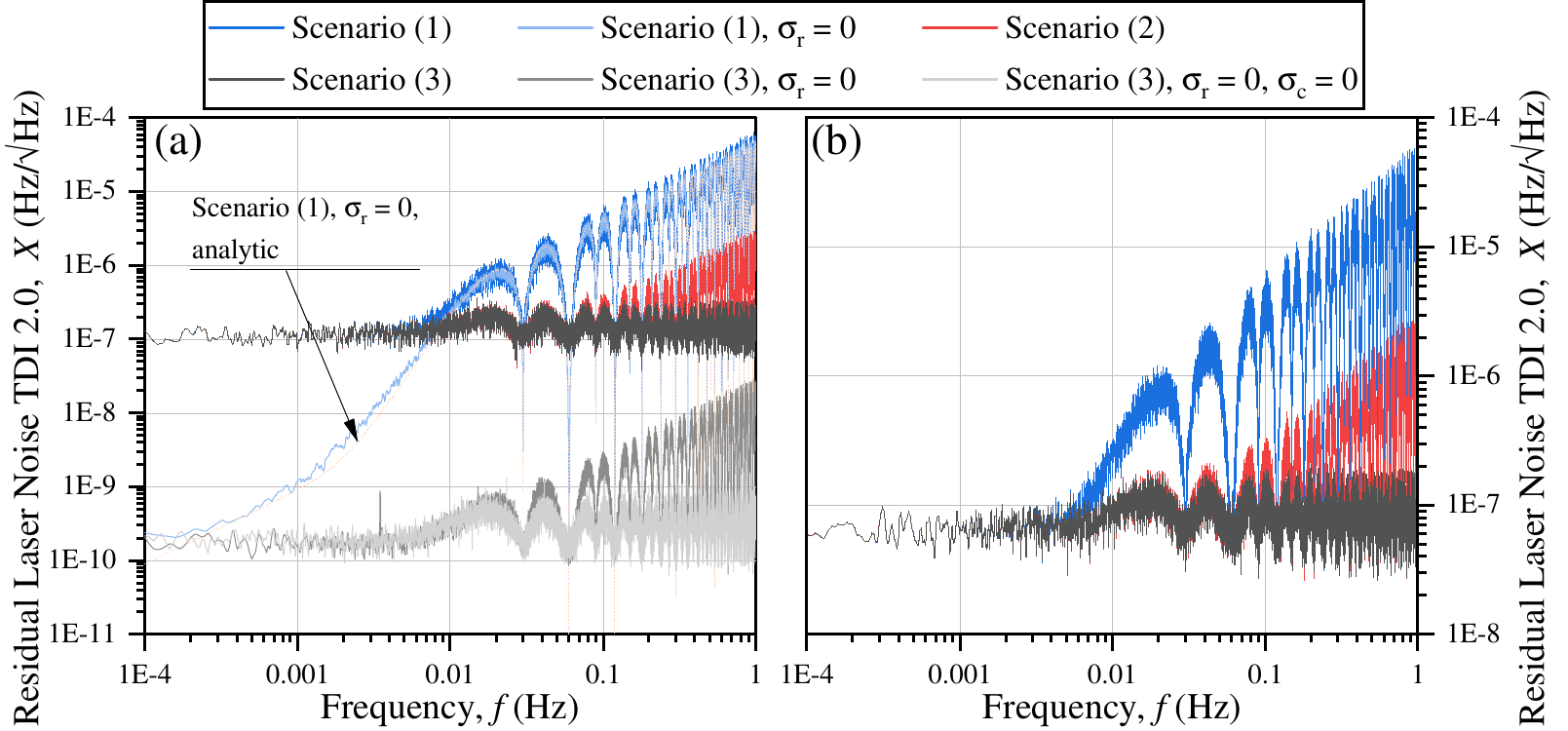}\caption{Residual laser noise in the TDI $X$ variable for 3 different scenarios.
Panels (a) and (b) illustrate the results for the baseline ranging
scheme and an improved ranging scheme, respectively. \label{fig:TDI-Simulation}}
\end{figure*}

The resulting residual laser noise for the TDI $X$ variable is depicted
in Fig. \ref{fig:TDI-Simulation}(a). The gray curves consider a calibration
of the front-end delays in space (scenario 3) as delineated in section
\ref{subsec:Calibration-Measurement}. Thereby, the light gray curve
is free of any ranging or calibration errors, exhibiting a nearly
flat noise spectrum of $\approx2^{-10}\,\text{Hz}/\sqrt{\text{Hz}}$.
The (mid) gray curve is affected by calibration errors leading to
an increase in residual laser noise at frequencies $f>10$ mHz. The
maximum increase of the laser noise is at the high frequencies of
the measurement bandwidth by around one order of magnitude. Finally,
the dark gray curve exhibits calibration and ranging errors. Here,
the residual noise is increased by around three orders of magnitude
over the whole measurement bandwidth, compared to the previous cases.
Importantly, this exposes the ranging errors as the dominating noise
source, if the calibration scheme is considered. In contrast, the
red curve considers only a single, on-ground calibration of front-end
and modulation delays, and inevitably neglects fluctuations in these
delays (scenario 2). This leads to an increased laser noise resulting
from the front-end and modulation delays manifesting as an increased
residual noise at higher frequencies ($f>0.1$ Hz) by roughly one
order of magnitude compared to scenario (3). Finally, the blue curves
do not consider any mitigation of the front-end and modulation delays
(scenario 1). Towards low frequencies, the dark blue curve approaches
the noise level where a calibration is considered (overlapping with
dark gray curve), but the noise level increases by two orders of magnitude
towards higher frequencies compared to scenario (3). As a reference,
the light-blue curve is free of any ranging error. The laser frequency
noise is minimal at low frequencies, where front-end and modulation
delays have little effect and the associated noise is negligible.
At higher frequencies, the noise level strongly increases and exhibits
a similar noise level as the dark blue curve, exposing modulation
and front-end delays as dominant noise sources. This behavior is
accurately described by the analytical solution represented by the
orange dotted curve.

We note that the residual laser noise is dominated by the ranging
errors when taking into account the proposed calibration scheme. Improvement
of the ranging performance can be accomplished via different approaches:

(1) Time-delay interferometry ranging (TDIR) was initially introduced
in \cite{tdiranging1} and aims at minimizing the rms power within
specific TDI combinations to derive precise estimates of inter-S/C
time delays. The estimates prove to be sufficiently accurate for effectively
suppressing laser noise to a degree significantly lower than the secondary
sources of noise. An alternative application of TDIR to estimate common-mode
delays is given in \cite{Houba2023}. TDIR as a valuable crosscheck
mechanism for evaluating residual PRN ranging offsets is described
in \cite{reinhardt2023ranging}.

(2) An optimization of the existing principle is obtained most easily
by improving the carrier-to-noise ratio of the incoming signal of
the DLL. The major noise contribution is the interfering code of the
local laser \cite{sutton2010laser}. This noise source can be mitigated
by wiping off the interfering code at the beginning of the DLL \cite{Sutton_2013}.
This technique relies on two conditions. First of all, the delay of
the local code is well-known based on the calibration measurement
of the long-arm interferometer. Secondly, distortion of the chips
due to the PLL processing can be incorporated via on-ground simulations
\cite{EuringerTIM2023,sutton2010laser,Delgado2012}. We note that
the presence of remote and local code does not change this processing
as long as the PLL is working in the linear domain. Assuming perfect
wiping of the local reference code, the ranging performance can be
improved by around a factor of two, see last entry of Tab. \ref{tab:RMS-calibration-error}.
Interference code wiping has been experimentally demonstrated in \cite{Sutton_2013},
where suppression of the interfering chip sequence by a factor of
$\approx$ 8 could be achieved, resulting in an rms ranging error
of 0.06 m over a 0.5 Hz signal bandwidth.\\
For the improved ranging performance as stated in Tab. \ref{tab:RMS-calibration-error},
the resulting residual laser noise of the TDI $X$ variable is depicted
in Fig. \ref{fig:TDI-Simulation}(b). While the slope of the 3 scenarios
remains identical to Fig. \ref{fig:TDI-Simulation}(a), the noise
level is reduced by a factor of two at low frequencies. At higher
frequencies, noise levels increase due to the uncompensated front-end
and modulation delays for scenarios (1) and (2). This emphasizes the
importance of re-calibration, as considered in scenario (3), which
enables a reduction of the noise level at higher frequencies of the
LISA measurement band. In fact, even much smaller ranging errors below
1 mm can be achieved by combining the ranging output with orbit prediction,
for example using a Kalman filter as was used in \cite{reinhardt2023ranging}.
This further underscores the need for accurate calibration and compensation
of front-end and modulation delays.

\section{Conclusion}

In this paper, we introduced a novel method for calibrating front-end
and modulation delays in both, ground and space settings, allowing
for the compensation of these delays in the TDI Michelson variables.
\\
Based on a thorough identification of delays affecting ranging and
phase measurements in all interferometers, delays were categorized
into known and unknown delays, with known delays being easily determined
through simulations or geometric path-length measurements. The unknown
delays, specifically front-end and modulation delays, are difficult
to mitigate because they require intricate calibrations and are expected
to change in orbit.

To address this problem, a set of calibration measurements was proposed
that relies on the fact that carrier phase and code chips are expected
to experience very similar delays. These measurements, integrated
with minimal functional additions to the receiver architecture, enable
to express front-end and modulation delays in terms of one reference
delay per S/C, resulting in three reference delays for the constellation.
By incorporating this calibration set into the data processing pipeline,
we demonstrated the separation of the three remaining reference delays
in the three TDI Michelson variables. This led to the mitigation of
front-end and modulation delays in the TDI Michelson variables.

Finally, the performance of the calibration measurements was determined
based on the LISA receiver architecture and signal parameters. These
results allowed a validation of the proposed calibration scheme using
numerical simulations. Three scenarios were set up to evaluate the
residual laser noise in the TDI variables for front-end and modulation
delays being uncompensated, compensated via on-ground calibration,
and compensated via additional re-calibration in space. In the absence
of a calibration scheme, there is a significant increase in residual
laser noise at frequencies above 10 mHz. On-ground calibration reduces
the residual laser noise by one order of magnitude. Moreover, the
increase in residual laser noise due to front-end and modulation delays
shifts to frequencies above 100 mHz. Re-calibration in space eliminates
any performance degradation caused by front-end and modulation delays.
In this case, performance is limited by the ranging error, where improvement
techniques have been delineated.

In conclusion, our findings underscore the critical importance of
accurate delay calibration and the continuous adjustment of calibration
parameters throughout the operational lifetime of LISA.
\begin{acknowledgments}
The authors thank R. Flatscher, T. Ziegler, S. Delchambre, P. Gath
and P. Voigt for their support and fruitful discussions. \\
This work was supported by funding from the Max-Planck-Institut für
Gravitationsphysik (Albert-Einstein-Institut), based on a grant by
the Deutsches Zentrum für Luft- und Raumfahrt (DLR). The work was
supported by the Bundesministerium für Wirtschaft und Klimaschutz
based on a resolution of the German Bundestag (Project Ref. Number
50 OQ 1801).
\end{acknowledgments}

\appendix

\section{Biased Dynamic Time-Shift Operator\label{sec:Appendix:-Biased-Dynamic}}

In the following, we will derive an expression for the nested application
of the dynamic time-shift operator considered in TDI 2.0 in the presence
of front-end and modulation delays.

In TDI 1.0, the distances between S/C are presumed to remain constant
over time. TDI 2.0 departs from this assumption by taking into account
the temporal fluctuations in these distances. This is incorporated
based on a dynamic time-shift operator $\mathcal{D}$ commonly defined
as \cite{OttoThesis2015}
\begin{equation}
\mathcal{D}_{ba}^{\text{ideal}}f(t):=f\left(t-\tau_{\text{R},ba}(t)\right),\label{eq:idela_dyn}
\end{equation}
with $a,b\in\{1,2,3\}$ and $a\neq b$. Hereby, $\tau_{\text{R},ba}(t)$
denotes the time-dependent ranging delay among S/C $a$ and S/C $b$
at time $t$. The superscript ``ideal'' in the dynamic time-shift
operator denotes that definition \ref{eq:idela_dyn} considers the
S/C as point masses. Nested application results in
\begin{align}
\mathcal{D}_{cb}^{\text{ideal}}\mathcal{D}_{ba}^{\text{ideal}}f(t) & =\mathcal{D}_{cb}^{\text{ideal}}f\left(t-\tau_{\text{R},ba}(t)\right),\nonumber \\
 & =f\left(t-\tau_{\text{R},cb}(t)-\tau_{\text{R},ba}(t-\tau_{\text{R},cb}(t))\right),\label{eq:idela_dyn_nested}
\end{align}
and similar for $n$-fold delays. Indices in Eq. \ref{eq:idela_dyn_nested}
are defined as $a,b,c\in\{1,2,3\}$ and $a\neq b\neq c$.

In analogous manner to the biased time-shift operator of Eq. \ref{eq:D_biased_ba},
we define the biased dynamic time-shift operator as
\begin{align}
\mathcal{D}_{ba}^{\text{bias}}f(t) & :=f(t-\tau_{ba}(t)+m_{\text{Mod},ab}+m_{\text{FE},ba,\text{L}}).\label{eq:def_biased_dyn_D}
\end{align}
Front-end and modulation delays and consequently the measurement
parameters $m_{\text{Mod},ab}$ and $m_{\text{FE},ba,\text{L}}$ are
assumed to be slowly varying on the timescale of weeks or months as
described in section \ref{sec:Calibration-performance}. Moreover,
the variation of these delays is expected to be on the order of $\pm$5
ns. The time-shifts induced by the ideal and the biased time-shift
operator are on the order of tens of seconds \cite{EESA}. On this
time scale the variation of the parameter $m_{\text{Mod},ab}$ as
well as $m_{\text{FE},ba,\text{L}}$ is on the order of $10^{-13}$
s, which is well below the stated calibration and ranging accuracy,
cf. Table \ref{tab:RMS-calibration-error}. Therefore, the time dependence
is omitted for the measurement parameters $m_{\text{Mod},ab}$ and
$m_{\text{FE},ba,\text{L}}$ in definition \ref{eq:def_biased_dyn_D}
and will be omitted for front-end and modulation delays in the ongoing
discussion.\\
Nested application of the biased dynamic time-shift operator results
in

\begin{align}
 & \mathcal{D}_{cb}^{\text{bias}}\mathcal{D}_{ba}^{\text{bias}}f(t-\tau_{\text{FE},ac,\text{L}})\nonumber \\
 & =\mathcal{D}_{cb}^{\text{bias}}f(t-\tau_{ba}(t)+m_{\text{Mod},ab}+m_{\text{FE},ba,\text{L}}-\tau_{\text{FE},ac,\text{L}})\nonumber \\
 & =\mathcal{D}_{cb}^{\text{bias}}f\left(t-(\tau_{\text{Mod},ab}+\tau_{\text{R},ba}(t)+\tau_{\text{FE},ba,\text{L}})\right.\nonumber \\
 & \left.+m_{\text{Mod},ab}+m_{\text{FE},ba,\text{L}}-\tau_{\text{FE},ac,\text{L}}\right)\nonumber \\
 & =\mathcal{D}_{cb}^{\text{bias}}f\left(t-m_{\text{Mod},ab}+\tau_{\text{FE},ac,\text{L}}-\tau_{\text{R},ba}(t)\right.\nonumber \\
 & \left.-m_{\text{FE},ba,\text{L}}-\tau_{\text{FE,}bd,\text{L}}+m_{\text{Mod},ab}+m_{\text{FE},ba,\text{L}}-\tau_{\text{FE},ac,\text{L}}\right)\nonumber \\
 & =\mathcal{D}_{cb}^{\text{bias}}f(t-\tau_{\text{R},ba}(t)-\tau_{\text{FE,}bd,\text{L}})\nonumber \\
 & =f\left(t-\tau_{cb}(t)\right.\nonumber \\
 & -\tau_{\text{R},ba}(t-\tau_{cb}(t)+m_{\text{Mod},bc}+m_{\text{FE},cb,\text{L}})\nonumber \\
 & \left.+m_{\text{Mod},bc}+m_{\text{FE},cb,\text{L}}-\tau_{\text{FE,}bd,\text{L}}\right)\nonumber \\
 & =f\left(t-(\tau_{\text{Mod},bc}+\tau_{\text{R},cb}(t)+\tau_{\text{FE},cb,\text{L}})\right.\nonumber \\
 & -\tau_{\text{R},ba}(t-(\tau_{\text{Mod},bc}+\tau_{\text{R},cb}(t)+\tau_{\text{FE},cb,\text{L}})\nonumber \\
 & +m_{\text{Mod},bc}+m_{\text{FE},cb,\text{L}})\nonumber \\
 & \left.+m_{\text{Mod},bc}+m_{\text{FE},cb,\text{L}}-\tau_{\text{FE,}bd,\text{L}}\right)\nonumber \\
 & =f(t-\tau_{\text{R},ba}(t-\tau_{\text{R},cb}(t)-\tau_{\text{FE},ce,\text{L}})-\tau_{\text{R},cb}(t)-\tau_{\text{FE},ce,\text{L}}),\label{eq:dyn_delay_nested}
\end{align}
where we extensively employed the calibration scheme provided in Eq.
\ref{eq:calibration_matrix_xy-1} and indices are defined as $a,b,c,d,e\in\{1,2,3\}$
and $a\neq b\neq c,a\neq c\neq e,b\neq d$. To evaluate the additional
shift by $\tau_{\text{FE},ce,\text{L}}$ in the argument of $\tau_{\text{R},ba}$
we perform a first-order Taylor expansion of $\tau_{\text{R},ba}$
\begin{align}
\tau_{\text{R},ba}(t_{\text{ref}}-\tau_{\text{FE},ce,\text{L}}) & \approx\tau_{\text{R},ba}(t_{\text{ref}})-\dot{\tau}_{\text{R},ba}(t_{\text{ref}})\tau_{\text{FE},ce,\text{L}}\nonumber \\
 & \approx\tau_{\text{R},ba}(t_{\text{ref}})-\dfrac{\dot{L}_{\text{R},ba}(t_{\text{ref}})}{c}\tau_{\text{FE},ce,\text{L}}.\label{eq:TaylorExp}
\end{align}
Hereby, $t_{\text{ref}}$ represents an arbitrary reference point
in time and $c$ the speed of light. The parameter $\dot{L}_{\text{R},ba}(t_{\text{ref}})$
denotes the relative velocity along the line of sight among S/C $a$
and S/C $b$ which is around 15 m/s \cite{OttoThesis2015}. As a result,
$\tau_{\text{FE},ce,\text{L}}$ is weighted by the factor of $\dot{L}_{\text{R},ba}(t_{\text{ref}})/c\approx10^{-8}$
in Eq. \ref{eq:TaylorExp}. Inserting Eq. \ref{eq:TaylorExp} in Eq.
\ref{eq:dyn_delay_nested} results in
\begin{align}
\mathcal{D}_{cb}^{\text{bias}} & \mathcal{D}_{ba}^{\text{bias}}f(t-\tau_{\text{FE},ac,\text{L}})\nonumber \\
 & =f\left(t-\tau_{\text{R},ba}\left(t-\tau_{\text{R},cb}(t)\right)-\tau_{\text{R},cb}(t)\right.\nonumber \\
 & \left.-\underbrace{\dfrac{\dot{L}_{\text{R},ba}\left(t-\tau_{\text{R},cb}(t)\right)}{c}\tau_{\text{FE},ce,\text{L}}}_{\approx10^{-15}}-\tau_{\text{FE},ce,\text{L}}\right)\nonumber \\
 & \approx f\left(t-\tau_{\text{R},ba}\left(t-\tau_{\text{R},cb}(t)\right)-\tau_{\text{R},cb}(t)-\tau_{\text{FE},ce,\text{L}}\right).\label{eq:D_bias_var}
\end{align}
In the last line, we neglected the additional contribution of $\tau_{\text{FE},ce,\text{L}}$
which is on the order of femtoseconds and thus well below the stated
calibration and ranging accuracy, cf. Table \ref{tab:RMS-calibration-error}.
Finally, combining Eq. \ref{eq:idela_dyn_nested} and Eq. \ref{eq:D_bias_var},
reveals the relation between nested application of the biased and
the ideal dynamic time-shift operator
\begin{align}
\mathcal{D}_{cb}^{\text{bias}}\mathcal{D}_{ba}^{\text{bias}}f(t-\tau_{\text{FE},ac,\text{L}}) & \approx\mathcal{D}_{cb}^{\text{ideal}}\mathcal{D}_{ba}^{\text{ideal}}f(t-\tau_{\text{FE},ce,\text{L}}).\label{eq:nested_dyn_D}
\end{align}
Comparison of Eq. \ref{eq:nested_dyn_D} with Eq. \ref{eq:Recursive applicaiton biased time shift}
manifests that the relation between the nested application of the
biased and the ideal time-shift operator is identical for the dynamic
time-shift operator considered in TDI 2.0 and the static one considered
in TDI 1.0. 

\bibliography{ref_publication_2023_1_manuscript}

\end{document}